\documentclass[acmsmall,manuscript,screen]{acmart}
\AtBeginDocument{%
  \providecommand\BibTeX{{%
    \normalfont B\kern-0.5em{\scshape i\kern-0.25em b}\kern-0.8em\TeX}}}

\setcopyright{acmcopyright}
\copyrightyear{2024}
\acmYear{2024}
\acmDOI{XXXXXXX.XXXXXXX}

\begin{document}

%%
%% The "title" command has an optional parameter,
%% allowing the author to define a "short title" to be used in page headers.
\title{Reclaiming Power over AI: Equipping Queer Teens as AI Designers for HIV Prevention}

%%
%% The "author" command and its associated commands are used to define
%% the authors and their affiliations.
%% Of note is the shared affiliation of the first two authors, and the
%% "authornote" and "authornotemark" commands
%% used to denote shared contribution to the research.
\author{William Liem}
\authornote{Corresponding author}
\email{william-liem@northwestern.edu}
\orcid{0000-0002-5873-6470}
\affiliation{%
  \institution{Northwestern University}
  \streetaddress{633 North St. Clair, Floor 20}
  \city{Chicago}
  \state{Illinois}
  \country{USA}
  \postcode{60611}
}

\author{Andrew Berry}
\affiliation{%
  \institution{Northwestern University}
  \city{Chicago}
  \country{USA}}
\email{andrew.berry@northwestern.edu}

\author{Kathryn Macapagal}
\affiliation{%
  \institution{Northwestern University}
  \city{Chicago}
  \country{USA}}
\email{kathryn.macapgal@northwestern.edu}

%%
%% By default, the full list of authors will be used in the page
%% headers. Often, this list is too long, and will overlap
%% other information printed in the page headers. This command allows
%% the author to define a more concise list
%% of authors' names for this purpose.
\renewcommand{\shortauthors}{Liem, et al.}

%%
%% The abstract is a short summary of the work to be presented in the
%% article.
\begin{abstract}
In this position paper, we explore the potential of generative AI (GenAI) tools in supporting HIV prevention initiatives among LGBTQ+ adolescents. GenAI offers opportunities to bridge information gaps and enhance healthcare access, yet it also risks exacerbating existing inequities through biased AI outputs reflecting heteronormative and cisnormative values. We advocate for the importance of queer adolescent-centered interventions, contend with the promise of GenAI tools while addressing concerns of bias, and position participatory frameworks for empowering queer youth in the design and development of AI tools. Viewing LGBTQ+ adolescents as designers, we propose a community-engaged approach to enable a group of queer teens with sexual health education expertise to design their own GenAI health tools. Through this collaborative effort, we put forward participatory ways to develop processes minimizing the potential iatrogenic harms of biased AI models, while harnessing AI benefits for LGBTQ+ teens. In this workshop, we offer specialized community-engaged knowledge in designing equitable AI tools to improve LGBTQ+ well-being.
\end{abstract}


%%
%% Keywords. The author(s) should pick words that accurately describe
%% the work being presented. Separate the keywords with commas.

%%
%% This command processes the author and affiliation and title
%% information and builds the first part of the formatted document.
\maketitle

\section{Introduction}
The advent of GenAI introduces immense promise in personalized healthcare, however, biased AI models risk further marginalizing historically vulnerable populations such as the LGBTQ+ community. Particularly, LGBTQ+ adolescents encounter distinct health disparities, highlighted by alarming rates of HIV infections and challenges in discreetly accessing HIV preventive services, often hindered by laws restricting minors from self-consenting \cite{noauthor_minors_2022}. GenAI tools like ChatGPT have the potential to enable LGBTQ+ adolescents to seek sexual health information anonymously. However, in alignment with open problems for AI in designing for well-being, we grapple with the relevance and effectiveness of AI outputs in addressing such disparities. Given the complexity of personal health data and biases in AI models, addressing these issues emerges as critical focal points. Moreover, we argue that building LGBTQ+ adolescents' AI literacy and involving them in co-design processes are essential for sustaining diversity, inclusivity, and adaptability in AI-driven healthcare designs.

Within this context, GenAI tools offer promising avenues for bridging informational gaps and enhancing healthcare access. However, inherent biases within AI models can cause undue harm to queer users. For example, AI models are trained on cisnormative and heteronormative datasets \cite{costanza2018design} and are capable of misgendering users \cite{keyes_misgendering_2018}. This can exacerbate emotional distress among transgender users who have historically faced systemic identity neglect \cite{scheuerman_how_2019}. In this position paper, we advocate for an approach aimed at co-developing safeguards against potential AI harm to LGBTQ+ adolescents and exploring potential opportunities for GenAI tools in addressing teen sexual health education. We highlight the need for adolescent-centric HIV prevention strategies among LGBTQ+ adolescents (section ~\ref{background1}), survey the potential promise of ChatGPT for personalized LGBTQ+ inclusive healthcare (section ~\ref{background2}), and propose a combinatory approach to co-designing AI tools that mitigate potential harms of biased models (section ~\ref{background3}). We conclude with a forthcoming research plan that responds to the growing need for involving LGBTQ+ adolescents in the interrogation and development of AI tools. We anticipate this work to demonstrate the potential for AI scientists, policymakers, and practitioners to work with disenfranchised queer populations in creating more equitable AI health tools. 

\section{Background}

\subsection{Co-Designing HIV Prevention for LGBTQ+ Youth}
\label{background1}

In 2021, people aged 13 to 24 accounted for almost 20 percent of new HIV infections, \cite{noauthor_hiv_2024} with the COVID-19 pandemic significantly disrupting the availability of HIV health services, \cite{rick_impact_2022} including pre-exposure prophylaxis (PrEP), a once-daily medication with 99\% effectiveness at preventing HIV transmission \cite{cdc_prep_2022}. This trend underscores the pressing need to address HIV prevention and education strategies, particularly among this vulnerable population. Higher HIV incidence rates may be attributed to the discontinuation of PrEP use by 1 in 7 young men who have sex with men aged 17-24 \cite{hong_prep_2022}. Younger people are particularly vulnerable, as they are more likely to discontinue PrEP \cite{chow_brief_2021} and less likely to reinitiate PrEP \cite{rao_persistence_2022}. This discontinuation trend is emblematic of broader challenges within the LGBTQ+ youth community, which faces a disproportionately higher burden of HIV compared to their heterosexual peers \cite{ocfemia_hiv_2018}. Multiple studies reveal that only half of LGBTQ+ teens know about PrEP’s existence and less than 3 percent have ever used it \cite{dunville_awareness_2021, macapagal_perspectives_2021, matson_awareness_2021, moskowitz_prep_2021, moskowitz_what_2020, thoma_brief_2018}. In response to these challenges, innovative approaches such as co-author Dr. Macapagal’s Teen Health Lab’s work in developing adolescent-centered digital HIV prevention programs and social marketing campaigns have emerged to empower youth in HIV prevention, advocacy, and education \cite{macapagal_64_2022}. By involving youth in the creation of their own health promotion efforts, participatory initiatives like the Teen Health Lab represent a promising model and ethos for reducing HIV incidence among young people by cultivating a sense of ownership and agency in HIV prevention strategies.

However, systemic barriers still plague LGBTQ+ youth regarding PrEP access and information. Parents are key gatekeepers for PrEP uptake as LGBTQ+ youth rely on them for navigating health care and associated costs \cite{mckay_parent-adolescent_2020}, however, there is a risk of disclosing their sexual identity in the process of exploring PrEP \cite{fisher_free_2018, moskowitz_what_2020}. Recent work examining experiences of PrEP uptake among 100 adolescent sexual minority men in the US calls for interventions to enhance adolescents’ self-efficacy skills and PrEP knowledge, such as sexual health care coaching to discuss PrEP uptake with parents or providers, enabling them to pursue PrEP independently if desired \cite{gordon_experiences_2024}. With 56\% of young people aged 18-24 using GenAI tools for information searching \cite{noauthor_digital_nodate},  there is a prime opportunity to explore the role of GenAI in addressing barriers to PrEP information seeking among LGBTQ+ youth. In the next section, we review innovative approaches to leverage GenAI tools for personalized healthcare, and contextualize it within an LGBTQ+ adolescent context.

\subsection{Generative AI for Personalized Healthcare}
\label{background2}
Recent studies have shown that AI chatbots potentially improve knowledge about and uptake of PrEP \cite{braddock_increasing_2023, hassani_potential_2022} and ChatGPT specifically shows some promise in counseling people living with HIV on treatment \cite{koh_role_2024}. Further, GenAI tools like ChatGPT are promising for delivering simplified health information \cite{ayre_new_2023, biswas_role_2023} and promoting health literacy \cite{coskun_integration_2024}, however, caution is needed to mitigate the potential harms of biased algorithms towards the queer community \cite{bragazzi_impact_2023,tomasev_fairness_2021,mcara-hunter_how_2024}. Datasets used to train chatbots and internet-based assistance are often riddled with gender biases \cite{gross_what_2023, mittermaier_bias_2023} such as the conflation of sex and gender, discourse of toxic masculinity, and erasure of non-binary identities \cite{seaborn_transcending_2023}, which can result in harmful outputs such as misgendering. This is particularly concerning given that many trans people already struggle with gender dysphoria, where gendered outputs may exacerbate emotional distress related to their gender identity or social experiences \cite{scheuerman_how_2019}. Further, these harms could cause people to disengage with GenAI tools completely, preventing them from reaping the potential benefits of accessible personalized health information.

Nascent work on curbing gendered biases in large language models has called for further work on creating “personalized guardrails” enabling end-users to design the boundaries of AI tools via prompt engineering as a means to combat gendered outputs of ChatGPT \cite{dwivedi_breaking_2023}. Scholars have sounded the alarm for engaging marginalized communities in co-designing AI tools that meet their needs \cite{bragazzi_impact_2023, tomasev_fairness_2021, mcara-hunter_how_2024, zytko_participatory_2022, buslon_raising_2023}. In particular, recommendations include promoting awareness of sex and gender biases in AI, advocating for fair AI-personalized medicine, and including diverse LGBTQ+ perspectives in either auditing or developing AI tools \cite{mahelona_openais_2023}.  We respond to these recommendations by surveying approaches that have engaged marginalized communities in the critique and interrogation of AI models in order to inform our research agenda.

\subsection{Community-Engaged AI Design}
\label{background3}
There is a pressing need to involve underrepresented and marginalized groups in designing and developing AI solutions to improve one’s well-being. Such involvement should build on the lived experiences of the user, incorporate the diversity of queer communities, and accommodate the multitude of considerations needed to design inclusive and equitable AI health tools. However, inviting everyday individuals to interrogate and make AI systems more equitable remains a perennial challenge, as it requires advanced technical knowledge that often acts as a barrier to participation \cite{sloane_participation_2022} GenAI tools like ChatGPT have opened a new format for an everyday audience to influence AI outputs where each individual user can set parameters and guidelines to shape the output. This presents a unique opportunity for community participation in the critique of AI tools, however, we need to be intentional with how participation is designed, as there are significant risks with “participation washing” where shallow forms of engagement inhibit the actual interrogation and redesign of these tools \cite{birhane_power_2022, agnew_technologies_2023}. We draw from two promising frameworks for designing AI tools with a non-technical audience: meta-design for resistance, and the PARticipatory Queer AI Research in Mental Health (PARQAIR-MH) protocol.

\subsubsection {Meta-design for resistance}
Resistance as characterized by Agnew et al.,\cite{agnew_technologies_2023}  is a process of adapting technologies to meet the community's needs. This framework addresses power imbalances inherent in technology-driven environments by empowering individuals and communities to oversee data collection, algorithmic decision-making, and AI model creation. This facilitates an environment where individuals can actively oppose undesirable AI applications within their lives and local settings. There are three levels of meta-design for resistance. 
\begin{enumerate}
    \item Adversarial Attacks and Privacy Enhancing Technologies as AI Defenses: The first level involves identifying tools to defend against AI harms, such as exploiting how biased AI models can cause unintentional harm, and creating guardrails to protect users from that.
    \item Collaboration for Resistance: The second level is collaboration for resistance. Sociotechnical in nature, it explores how AI defenses and collaborations can be designed together to empower resistance. An example is when TikTok users replace words that would trigger algorithms to flag their content as harmful, when they are trying to advocate for a social justice issue \cite{karizat_algorithmic_2021}.
    \item Communities of Resistance: Supportive networks that facilitate collaboration, knowledge sharing, and coordinated actions to resist AI harms, fostering empowerment and grassroots efforts to challenge and mitigate the negative impacts of AI technologies within marginalized communities. This bridges the gap between research and practical application, while also shaping broader cultural expectations of contestability and ownership in the face of data and AI-related challenges.
\end{enumerate}

Agnew et al., argue that central to this framework is the concept of designing AI tools that align with the values of the community. This is necessary because one goal of this framework is to enable end users to provide feedback and iterate so that if there is value drift, communities have the ability to realign the technology or stop its deployment. In order to achieve a consensus on these values to center in the design process, we draw from the PARticipatory Queer AI Research in Mental Health (PARQAIR-MH) protocol. 

\subsubsection{PARQAIR-MH}
This community-engaged Delphi process incorporates queer perspectives on AI design in mental health care \cite{joyce_protocol_2023}. This protocol aims to deliver a toolkit that will help ensure that the specific needs of the LGBTQ+ communities are accounted for in mental health applications of data-driven technologies. The protocol starts with a comprehensive literature review of the problem space, with the goal of sharing findings back to queer stakeholders to ground foundational knowledge of the AI problem space. Subsequently, through a series of Delphi rounds, the group refines a moral framework for how AI tools should be used and developed that meet the needs of their community. It assists researchers, healthcare organizations, and policymakers in deciding how to collect and use data on sensitive characteristics appropriately. 

We adapt this process to be inclusive, but not limited to, mental health care in AI. We believe this protocol embodies scientific rigor through a Delphi process for achieving a consensus across diverse perspectives, while also adhering to the central tenets of community-based participatory research. We conclude this position paper with a proposed research plan, one where we use the aforementioned frameworks to engage an adolescent queer community to be AI designers and have them explore concerns and opportunities as it relates to LGBTQ+ health. 

\section{Proposed Research Direction}

Building across our multidisciplinary team in digital sexual health, community-engagement, and values-aligned technology, we plan on engaging a Youth Advisory Council (YAC) managed by the Teen Health Lab (led by co-author Dr. Macapgal) housed within Northwestern’s Institute of Sexual and Gender Minority Health and Wellbeing. In the near term, we aim to grow a GenAI working group between our team and YAC. The YAC uses a private, teen-only Discord server facilitated by young LGBTQ research staff where LGBTQ high school students across the U.S. have offered meaningful input into four teen-centered digital and multimedia sexual health projects over the last year. This YAC has co-designed a text-messaging intervention aimed at promoting engagement in sexual healthcare among LGBTQ+ adolescents and offered feedback on an adolescent sexual consent intervention. We have worked with this group on co-designing PrEP4teens.com \cite{prep4teens}, a website for promoting and informing PrEP.

The YAC is representative of a wide variety of sexual, gender, and racial identities–members who are historically excluded from the participation of AI tools, which has been repeatedly cited to be a significant contributor to biased AI algorithms \cite{ayre_new_2023, biswas_role_2023, coskun_integration_2024}. Our research aims to leverage the vested interests of the YAC and create an opportunity for them to shape AI tools in the way they deem necessary for equitable HIV information seeking and dissemination. We plan on following an adapted version of the PARQAIR-MH protocol to build on our existing infrastructure and equip the YAC with a foundational knowledge of GenAI, informing them about documented use cases and harms towards the queer community. This will prepare them to explore what opportunities, concerns, and considerations must be made to leverage these tools for HIV prevention and information dissemination equitably, resulting in a moral framework for co-design. 

Following suit, we follow the meta-design for resistance framework where we will lead a series of co-design workshops and usability testing to identify requirements for a queer-centric AI health assistant on ChatGPT and conduct one-on-one usability tests to exploit unintended harms. Through this, we will present back findings to the YAC to elicit reactions and create a robust list of guidelines to safeguard queer adolescents who may use ChatGPT for HIV prevention information. We plan on finding opportunities with the YAC for dissemination channels that would promote AI literacy.

We believe this is a promising approach to participatory methods in designing personalized AI health tools. Through engaging a diverse cast of queer teens who are experts in sexual health, we can build interventions to minimize the potential iatrogenic harms of biased AI models. In this workshop, we are poised to contribute our expertise in designing with marginalized communities to cultivate the development of equitable AI tools. We anticipate that this project will significantly benefit from the insightful discussions facilitated by the one-day workshop on designing with AI for well-being. 

%%
%% The next two lines define the bibliography style to be used, and
%% the bibliography file.
\bibliographystyle{ACM-Reference-Format}
\bibliography{Workshop}

\end{document}